\documentclass[Journal]{IEEEtran}
\usepackage{enumerate}
\usepackage{siunitx}
\usepackage{import} 
\usepackage{enumitem}

\usepackage{indentfirst} 
\usepackage{standalone}
\usepackage[section]{placeins} 
\usepackage{pdflscape}
\usepackage{algorithm} 

\usepackage{amssymb} 

\usepackage{microtype} 

\usepackage{soul} 
\usepackage{mathtools}
\usepackage{graphicx}
\usepackage[hidelinks]{hyperref}
\usepackage{pgfplots}

\usepackage{subfloat} 

\usepackage[utf8]{inputenc}
\usepackage{amsmath} 
\usepackage[shellescape,latex]{gmp} 
\usepackage[space]{grffile}
\usepackage{xfrac}
\usepackage{tabularx}

\usepackage{cite}
\usepackage{adjustbox}

\usepackage{verbatim}
\usepackage{url}

\usetikzlibrary{shapes}

\usepackage{multicol, blindtext}    
\usepackage[]{color}
\usepackage{framed}
\usepackage{alltt}
\usepackage[T1]{fontenc}
\usepackage{array}
\usepackage{colortbl}
\usepackage{booktabs}
\usepackage{multirow}
\usepackage{eurosym}
\usepackage{listings}
\usepackage{ifthen}
\usepackage{pgfplotstable}
\usepackage{pgfcalendar}
\usepackage{pgfgantt}
\usepackage{tikz}
\usetikzlibrary{shapes.geometric, arrows}

\usepackage{longtable}
\usepackage{amsfonts}
\usepackage{subfigure}
\usepackage{float}

\usepackage{mathrsfs}
\usepackage{algpseudocode} 
\usepackage{xcolor}
\usepackage{textcomp}
\usepackage{xspace} 

\newcommand{\PhaseLine}[1]{%
  \Statex                    
  \noindent
  \hspace*{-\algorithmicindent} 
  \begin{tikzpicture}
    \draw[densely dashed] (0,0) -- (\linewidth,0);
    \node[fill=white] at (\linewidth/2,0) {\textit{#1}};
  \end{tikzpicture}
}

\makeatletter
\newcommand{\LongState}[1]{%
  \State\leavevmode
  \hspace*{0pt}%
  \parbox[t]{\dimexpr\linewidth-\ALG@thistlm\relax}{\raggedright\strut #1\strut}%
}
\newcommand{\LongStatex}[1]{%
  \Statex\hspace*{\ALG@thistlm}%
  \parbox[t]{\dimexpr\linewidth-\ALG@thistlm\relax}{\raggedright\strut #1\strut}%
}
\makeatother

\pgfplotsset{compat=1.16}

\usepackage{fmtcount} 

\usepackage[bottom]{footmisc}
\begin{document}

\title{Incremental DRL-Based Resource Management for Dynamic Network Slicing in an Urban-Wide Testbed}

\author{Haiyuan Li, Yuelin Liu, Hari Madhukumar, Amin Emami, Xueqing Zhou, \\ Yulei Wu, Xenofon Vasilakos, Shuangyi Yan, Dimitra Simeonidou

\thanks{Haiyuan Li, Yuelin Liu, Hari Madhukumar, Amin Emami, Xueqing Zhou, Yulei Wu, Xenofon Vasilakos, Shuangyi Yan, Dimitra Simeonidou are with the High Performance Networks Group, Smart Internet Lab, Department of Electrical and Electronic Engineering, University of Bristol, BS8 1QU, U.K. (e-mail: ocean.h.li@bristol.ac.uk). Haiyuan Li and Yuelin Liu are co-first authors.}%

\thanks{Manuscript received XXXX YY, ZZZZ; revised XXXX YY, ZZZZ.}
}

\markboth{\LaTeX\ Class Files,~Vol.~X, No.~Y, Month~202X}%
{Shell \MakeLowercase{\textit{et al.}}: Bare Demo of IEEEtran.cls for Journals}

\maketitle 

\begin{abstract}
Multi-access edge computing provides localized resources within mobile networks to address the requirements of emerging latency-sensitive and computing-intensive applications. At the edge, dynamic requests necessitate sophisticated resource management for adaptive network slicing. This involves optimizing resource allocations, scaling functions, and load balancing to utilize only essential resources under constrained network scenarios. However, existing solutions largely assume static slice counts, ignoring the re-optimization overhead associated with management algorithms when slices fluctuate. Moreover, many approaches rely on simplified energy models that overlook intertemporal resource scheduling and are predominantly evaluated through simulations, neglecting critical practical considerations. This paper presents an incremental cooperative Multi-Agent Deep Deterministic Policy Gradient (MADDPG) algorithm for resource management in dynamic edge slicing. The proposed approach optimizes long-term slicing benefits by reducing delay and energy consumption while minimizing retraining overhead in response to slice variations. Furthermore, we implement an urban-wide edge computing testbed based on OpenStack and Kubernetes to validate the algorithm's performance. Experimental results demonstrate that our incremental MADDPG method outperforms benchmark strategies in aggregated slicing utility and reduces training energy consumption by up to 50\% compared to the re-optimization approach.
\end{abstract}


\maketitle

\section{INTRODUCTION}
With the rapid development of the Internet of Things (IoT) and mobile networks, there has been an increasing demand for latency-sensitive and computing-intensive services and applications~\cite{abouaomar2021resource, premsankar2022energy}.
In response, Multi-access Edge Computing (MEC) networks bring computational and storage resources closer to edge devices. By handling data and requests near the end-users, MEC networks significantly reduce latency, optimize traffic management, and facilitate a context-aware user experience. 
On the other hand, network slicing serves as a vital mechanism in 5G and beyond, allowing network operators to create multiple logical networks from a single physical infrastructure catering to various user requirements. 
The integration of Multi-access Edge Computing (MEC) and network slicing enhances the capabilities of 5G networks~\cite{thantharate2023adaptive6g, jin2023network, domeke2022integration, afolabi2018network}.
Given the proximity of data processing in MEC, network slices can offer their services with minimal latency, supporting the realization of real-time applications.

Nonetheless, a rising demand from network slices intensifies the competition for the given MEC resources. Resource allocations for a single slice affect not only its own performance but also that of others, contemporary or future, that utilize the same infrastructure.
Consequently, the constrained resource capacity and interconnected relationships among network slices through the shared resource infrastructure present a complex scheduling challenge. This underscores the importance of efficient resource allocation and management to optimize network performance.

Extensive research has been conducted to develop management strategies that allocate shared resources among slices to meet the needs of various 5G applications. These works can be categorized based on employed techniques: optimization-based~\cite{d2020sl, liu2019direct, ma2020slicing}, game theory-based~\cite{caballero2019network, datar2021strategic, tran2020resource, fossati2020multi}, and Deep Reinforcement Learning (DRL)-based strategies~\cite{hurtado2022deep, vila2020novel, sun2019dynamic, wu2022heterogeneous, van2019optimal, qi2019deep}~\cite{su2019resource, d2020sl}. They explored resource allocation, performance optimization, and the dynamics of network slices in the context of edge computing, each offering unique insights into managing limited resources and optimizing network performance.
However, the current research highlights three critical gaps that need resolution.
\begin{enumerate}[leftmargin=*, label=-]
    \item Research in this area has primarily focused on resource management for a fixed set of network slices, disregarding the need for re-optimization arising from slice variations. Such variations can disrupt inter-slice relationships, diminish the effectiveness of pre-trained algorithms, and restrict the applicability of pre-trained models.
    
    \item Existing research on energy optimization in MEC networks often relies on oversimplified models that overlook real-world complexities. First, many approaches assume simplistic (e.g., linear) power–utilization relationships or treat task energy consumption as a fixed value \cite{zhou2020dynamic, akin2022greenslice}. Second, some studies focus on instantaneous power rather than total energy, neglecting concurrency effects and power fluctuations as multiple tasks share server resources and new tasks arrive and complete.
        
    \item While current strategies provide valuable theoretical insights, their validation is largely confined to simulated environments~\cite{d2020sl, liu2019direct, ma2020slicing, caballero2019network, datar2021strategic, tran2020resource, fossati2020multi, hurtado2022deep, vila2020novel, sun2019dynamic, wu2022heterogeneous, van2019optimal, qi2019deep, su2019resource},  overlooking practical realities such as prolonged training times, delayed network feedback, and interference from model training on energy evaluations. The effectiveness of existing solutions in real-world deployments remains uncertain.
\end{enumerate}

This paper addresses the resource management problem of network slicing in multi-access edge computing by formulating it as an evolving Markov Decision Process (MDP), where network slice variations introduce new MDP instances, each potentially with distinct state and action spaces. Within this framework, our objective is to maximize long-term benefits in terms of processing latency and energy consumption across MEC servers while adapting to varying network slice configurations with minimal re-optimization costs. The main contributions of this paper are summarized below.

\begin{itemize}[leftmargin=*]
    \item In a single MDP, we propose a fully cooperative Multi-Agent Deep Deterministic Policy Gradient (MADDPG) wherein a DRL agent is assigned to each network slice, sharing a common reward. This strategy can capture the distinct demands and resource competition among slices, and reduce the action space of the entire network by partitioning it among multiple agents.

    \item In response to the dynamic changes in slice numbers and evolving MDP, an incremental learning scheme is introduced to improve the generalization capability of the algorithm and eliminate the requirement for re-training DRL models from scratch by inheriting knowledge from previous models.

    \item In evaluating the energy consumption, a stress test on a bare-metal server is conducted to establish an accurate power model. Furthermore, a delayed reward mechanism is incorporated into the MADDPG model to capture the task-dependent complexities of power usage and the cumulative impact of previous, immediate, and future actions on the energy consumption of shared servers.
    
    \item We deployed a testbed across multiple locations and established a Kubernetes cluster for networking and slice management. This setup enables real-time monitoring and robust resource orchestration. Leveraging this testbed, we validated the practical feasibility of the proposed incremental MADDPG algorithm and its ability to operate under realistic network conditions.
\end{itemize}

In the testbed, we conduct a comprehensive evaluation against a series of five benchmarks, including (i) a random allocation and (ii) an over-allocation approach, (iii) a static slicing-based solution originally proposed in~\cite{caballero2019network} (iv) a centralized single agent DDPG, and (v) a distributed self-interested MADDPG. As performance highlights, we prove that solutions without scheduling management cannot yield reliable results due to inadequate long-term consideration. 
Additionally, a single-agent DRL-based approach is unable to fully capture the interactions among slices, while a distributed, self-interested multi-agent framework tends to prioritize individual agent interests, potentially limiting overall performance.
In comparison, MADDPG-based algorithms can adapt to the diversity of resource variation and slice requests and obtain the highest mean return.
More importantly, with the assistance of incremental learning, MADDPG greatly reduces the training time and energy consumption compared to retraining from scratch. Experimental measurements indicate that the incremental approach lowers training energy consumption by up to 50\%.

The remainder of this paper is organized as follows. Section~\ref{sec:related} discusses related work. Section~\ref{sec:System} presents our adapted network slicing scenario and associated optimization problem formulation. Section~\ref{sec:methodology} elaborates on the intricacies of our proposed incremental multi-agent DRL solution. Section~\ref{sec:evaluation} offers a detailed account of our performance evaluation study, leveraging city-wide urban testbed environments for evaluation testing and validation purposes. Finally, Section\ref{sec:conclusion} concludes the key findings and outlines potential directions for future research endeavors.

\section{RELATED WORK}
\label{sec:related}

A considerable body of work in the literature targets devising management strategies for network slicing. The objective is to address the allocation of shared resources among slices within edge computing networks, catering to the diverse requirements of 5G applications.

\vspace{-0.4cm}
\subsection{Single agent-based approaches}
Sun et al.~\cite{sun2019dynamic} proposed an autonomous virtual resource-slicing framework that dynamically reserves resources based on traffic ratios, refining the allocation through a deep Q-network (DQN) algorithm. However, their approach, which employs a single agent to manage policies over the network, suffers from significant limitations due to the exponentially expanding action space with network slices increasing. Their evaluation was restricted to scenarios with only two slices, suggesting potential challenges in achieving convergence and adapting to more complex environments. In contrast, Suh et al.~\cite{suh2022deep} employed the same DRL algorithm to determine resource allocation for MECs across multiple slices. To address the action space explosion, they introduced an action elimination mechanism that filters out resource allocation decisions that violate Quality of Service (QoS) requirements, thereby streamlining the action space. 
Nevertheless, these strategies struggle to adequately account for the diverse requirements of different slices and to address the complex resource interdependencies among interconnected slices.

\subsection{Collaborative model or multi-agent approaches}
To support a growing number of slices and balance the distinct requirements and resource competition among slices, Caballero et al.~\cite{caballero2019network} established a resource-sharing model between slices, conceptualizing it as a Fisher market. They achieved convergence in this game at a Nash equilibrium, where each slice benefits from sharing performance while maintaining the ability to customize its own. However, this work concentrates on instantaneous performance and neglects resource dynamics over time.
In contrast, Vila et al.~\cite{vila2020novel} manage long-term cumulative returns by modeling the problem as a Markov Decision Process (MDP). To solve this MDP and simulate resource-sharing relationships among slices, they devised a cooperative multi-agent DQN algorithm. In this framework, each slice is managed by an individual DRL agent, which determines its capacity shares while accounting for Service Level Agreement (SLA) requirements across agents. 
However, these approaches overlook the impact of network slice variations, which create dynamic environments. Their proposed strategy of directly adding or removing agents fails to account for how changes in agent numbers can destabilize and reduce the effectiveness of pre-learned policies.

\subsection{Incremental approaches for addressing dynamic environments}
In order to improve the generalization capability of the model, Wadhwania et al.~\cite{wadhwania2019policy} proposed a policy distillation strategy that combines actor policies by creating a new agent that minimizes the Kullback-Leibler divergence concerning all agents. However, distilling knowledge between the new agent and multiple previously trained agents often demands a large dataset and iterative refinement, making it a complex, time-consuming process that incurs significant learning overhead.

In comparison, Agarwal et al.~\cite{agarwal2019learning} employed a Graph Neural Network (GNN) framework to model multi-agent systems, wherein each agent is represented as a node, and the interactions between agents are depicted as links. Leveraging the flexibility of GNNs in accommodating varying numbers of nodes, their architecture is adept at adapting to changes in the number of agents within the system.
However, this solution fails to specify the impact of increasing agents on the neural network architecture of individual agents.

\subsection{Considering energy consumption alongside other objectives}
In these studies, the optimization objectives are chiefly identified as QoS and SLA in terms of throughput and latency. Conversely, energy consumption, a critical aspect of network operational costs, has received insufficient attention.
Aiming to minimize the costs associated with Virtual Network Function (VNF) deployment and energy consumption, Zhou et al.~\cite{zhou2020dynamic} proposed an approach that integrates a Holt-Winters prediction algorithm for traffic demand estimation, an adaptive scaling strategy for optimal VNF and resource allocation, and a proactive deployment algorithm for network slicing. It demonstrates enhanced resource utilization, cost efficiency, and energy savings in network operations. To optimize VNF placement in 5G networks for enhanced energy efficiency,  Akin~\cite{akin2022greenslice} introduces an ILP model that considerably reduces power consumption while meeting stringent QoS and security requirements.
However, the energy models presented in these works are simplistic. In \cite{akin2022greenslice}, the energy consumption of a VNF is set to a static empirical value. In \cite{zhou2020dynamic}, a linear relationship is assumed between server loading and power consumption. 
Although studies such as \cite{ale2021delay, mao2017survey} use a more precise, non-linear model of CPU load to power mapping, the interdependence among concurrent tasks was still overlooked.

\vspace{-0.2cm}
\section{SYSTEM MODEL AND PROBLEM FORMULATION}
\label{sec:System}
\begin{table}[t]
  \centering
  
  \caption{Summary of Notations}
  \scalebox{1.1}{
    \begin{tabular}{rll}
    \toprule
          & \multicolumn{1}{c}{Notation} & \multicolumn{1}{c}{Description} \\
    \midrule
    \multicolumn{1}{l}{Index} & $i$     & Network slice / DRL agent \\
          & $m$     & MEC server  \\
          & $l$     & Link  \\
          & $t$     & Time slot inside state $\psi$  \\
          & $\psi$  & Network state period (MDP instance)  \\
    \midrule
    \multicolumn{1}{l}{Set} 
          & $I_\psi$   & Active‐slice set in state $\psi$ \\
          & $M_i$     & Server set used by slice $i$  \\
          & $L_i$     & Link set used by slice $i$ \\
          & $T_\psi$   & Slot set during $\psi$ state (duration $\tau_\psi$) \\
          & $\mathcal{J}_{M\psi t}$ & Remaining CPU set of all MECs at $\psi t$\\
          & $\mathcal{B}_{L\psi t}$ & Remaining BW set of all links at $\psi t$ \\
          & $\mathcal{G}_{I_\psi \psi t}$ & CPU‐demand set of all slices at $\psi t$ \\
          & $\mathcal{D}_{I_\psi \psi t}$ & Workload‐size set of all slices at $\psi t$ \\
    \midrule
    \multicolumn{1}{l}{Parameter} 
          & $J_m^{'}$     & CPU‐core capacity of server $m$ \\
          & $B_l^{'}$     & Bandwidth capacity of link $l$ \\
          & $G_{im\psi t}$ & Computing demands of $i$ on $m$ at $\psi t$ \\
          & $D_{il\psi t}$     & Workload of $i$ on $l$ at $\psi t$ \\
          & $\lambda_i$    & Latency preference weights \\
          & $\rho_i$     & Energy preference weights \\
          & $\gamma$     & DRL discount factor \\
    \midrule
    \multicolumn{1}{l}{Variable}
          & $J_{m\psi t}$ & Remaining CPU of $m$ at $\psi t$ \\
          & $B_{l\psi t}$ & Remaining BW of $l$ at $\psi t$ \\   
          & $g_{im\psi t}$     & CPU cores allocated to $i$ on $m$ at $\psi t$  \\
          & $d_{il\psi t}$     & Bandwidth allocated to $i$ on $l$ at $\psi t$  \\
          & $U_{m t}$     & Utilization ratio of server $m$ at $t$ \\
          & $W_{i\psi t}$     & End-to-end latency of slice $i$ at $\psi t$\\
          & $E_{i\psi t}$     & Energy consumed by slice $i$ at $\psi t$\\
          & $a_{im\psi t}$  & Actor output (CPU) on $m$ of $i$ at $\psi t$\\
          & $a_{il\psi t}$  & Actor output (BW) on $l$ of $i$ at $\psi t$\\
          & $r_{i\psi t}$     & Reward of slice $i$ at $\psi t$\\
          & $\eta_i$     & Policy-inheritance weight of slice $i$ \\
    \bottomrule
    \end{tabular}
    }
  \label{table:notations}
  \vspace{-0.3cm}
\end{table}%

\begin{figure}[t]
    \centering
    \vspace{0cm}
            \includegraphics[width=0.95\linewidth]{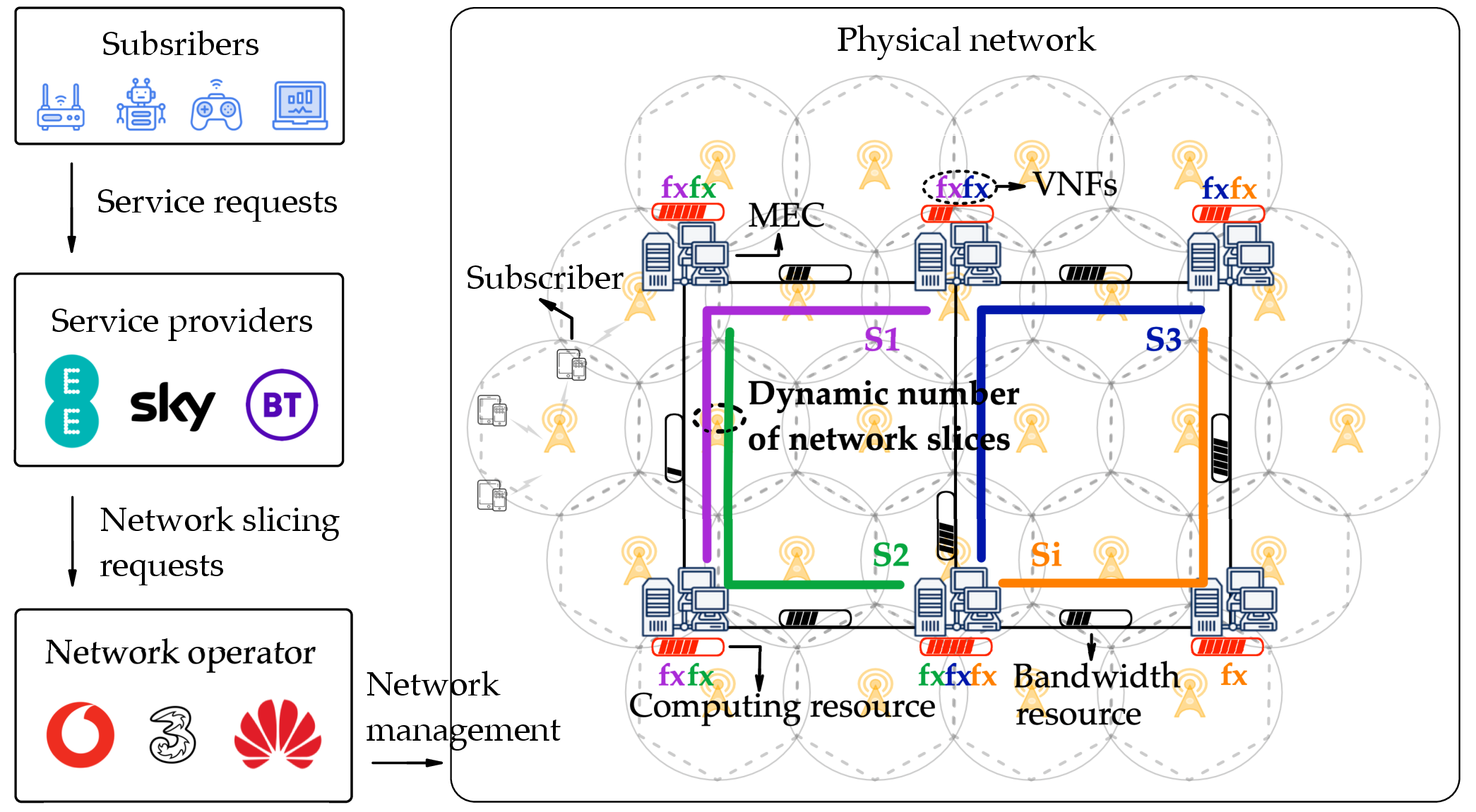}
    \caption{Network slicing in MEC networks with constrained computing and bandwidth resources.}
    \label{net_arch}
    \vspace{-0.3cm}
\end{figure}

An illustration of a multi-server edge computing network is presented in Figure~\ref{net_arch}. The principal entities within this network include MEC servers, links, subscribers, service providers, and a network slice operator. In this setup, subscribers request specific services or resources from service providers, such as video streaming, voice calls, internet connectivity, or other network services. Subscriber requests are processed as computational tasks executed through VNFs. VNFs represent essential network functions deployed within 5G networks, each characterised by specific resource demands on MEC servers. The execution duration of these VNFs explicitly depends on the computational resources and bandwidth allocated, directly impacting the latency experienced by users according to the size of their workloads.

An illustration of a multi-server edge computing network is presented in Figure~\ref{net_arch}. The principal entities within this network include MEC servers, links, subscribers, service providers, and a network slice operator. In this setup, subscribers request specific services or resources from vertical service providers, such as video streaming, voice calls, internet connectivity, or other network services. In this work, subscriber requests are abstracted as computational tasks that are executed through a set of VNFs with latency and energy consumption preferences. Rather than explicitly modeling specific 5G network functions (e.g., Access and Mobility Management Function (AMF), Session Management Function (SMF), User Plan Function (UPF)), we consider a generic collection of VNFs that impose heterogeneous computational and bandwidth demands on MEC servers. The execution time of a VNF depends on the allocated resources, which in turn affects the overall task latency experienced by users based on their workload size. It is important to note that we focus on task-level latency without considering finer granularity level latency predictions, such as at the level of streams or packets.

Before forwarding these requests to the network operator, each service provider enforces admission control policies to uphold SLAs and prevent system overload. Ultimately, the network operator is responsible for managing and optimizing the underlying MEC resources allocated to the VNFs of the admitted requests~\cite{li2017network, alliance2016description}.
In the system model, subscriber requests within each time interval represent aggregated demands for each slice type, capturing the accumulation of tasks arriving within that interval. All admitted requests from the previous time slot are processed at the beginning of the current slot. Therefore, no explicit queuing model is employed. For ease of reference, important notations used throughout this paper are summarized in Table I.

\subsection{Latency}
Within the MEC network, the overall latency $W_{it}$ for slice $i$ at time slot $t$ encompasses both computing latency $w_{ibt}$ and transmission delay $w_{iat}$. This metric influences both the duration of task processing and the time until the allocated resource for that request is freed. The expression for this latency is given by:
\begin{equation}
W_{it} = w_{iat} + w_{ibt}
\label{1}
\end{equation}

The VNF of each network slice undergoes processing on multiple MECs, the overall computing latency is contingent on the maximum delay among all MECs utilized by $i$, denoted as $M_{i}$. The computing latency $w_{ibt}$ can be expressed as:
\begin{equation}
w_{ibt} = \max_{m\in M_{i}} w_{imbt}
\label{2}
\end{equation}
where $w_{imbt}$ denotes the computing latency on MEC $m$, which can be calculated by the division of required computing resource $G_{imt}$ for VNFs on $m$ and allocated resource $g_{imt}$ 
\begin{equation}
    w_{imbt} = G_{imt} / g_{imt}
    \label{3}
\end{equation}
In addition, based on data rate $p_{ilt}$ on link $l$ and workload size on $l$, $D_{ilt}$, transmission latency of slice $i$ across multiple MECs, $w_{iat}$ can be determined as:
\begin{equation} 
w_{iat} = \sum_{l}^{L_{i}} D_{ilt} / p_{ilt}
\label{4}
\end{equation}
where $L_{i}$ represents the links passed by slice $i$. 
According to~\cite{schwartz2005mobile}, data rate $p_{ilt}$ on link $l$ can be simplified to:
\begin{equation}
p_{ilt} = d_{ilt} \log_{2} (1 + N)
\label{5}
\end{equation}
where $N$ and $d_{ilt}$ state for signal-to-noise ratio and allocated bandwidth on link $l$ to slice $i$ at time slot $t$~\cite{kiran2020vnf}, respectively.

\begin{figure}[t]
    \centering
    \includegraphics[width=0.95\linewidth]{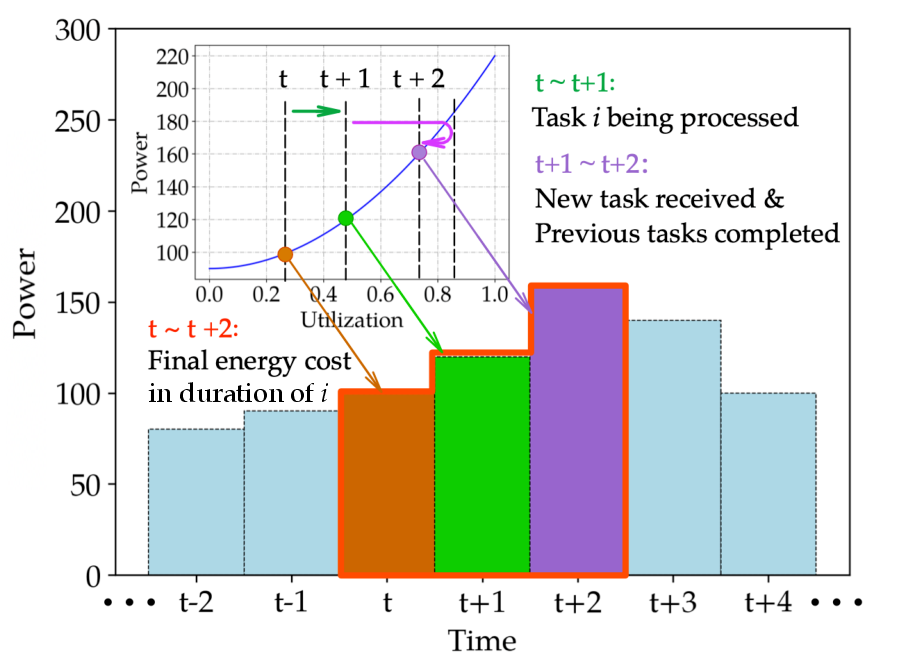}
    \caption{Dynamic power consumption of a MEC server caused by the resource allocation of previous and future requests. The top left corner showcases an example of energy consumption by a task, highlighting its co-dependence on past or future issued tasks besides its own resource allocation.}
    \label{energy_dynamics}
\end{figure}

\begin{figure}[!t]
    \centering
    \setlength{\belowcaptionskip}{0cm}
            \includegraphics[width=1\linewidth]{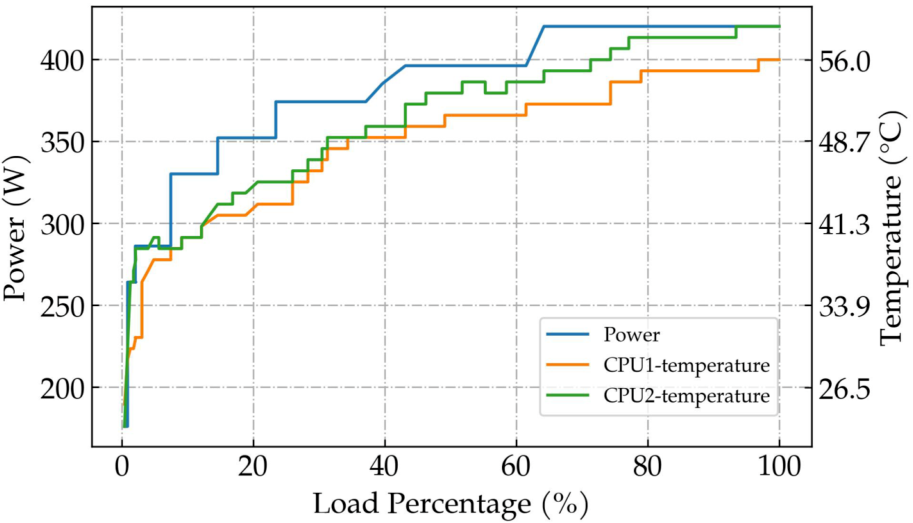}
    \caption{The power consumption and temperature of a Dell PowerEdge R760 bare-metal MEC server vs. CPU load percentages.}
    \label{powerstress}
\end{figure}

\subsection{Energy}
The energy consumption of each network slice is composed of contributions from multiple VNFs. For each VNF, its power consumption at a given time slot $t$ is not solely determined by the resource allocation at that particular moment but also by previous and future tasks and their corresponding resource allocation decisions.
As illustrated in the top-left corner of Figure~\ref{energy_dynamics}, consider a VNF running on a MEC server whose computation spans two consecutive time slots. In the subsequent time slots, new tasks may arrive while older tasks may be completed, thereby affecting the overall resource utilization and influencing the energy consumption of this VNF. Therefore, to capture the cumulative energy consumed by the MEC servers allocated to the slice $i$, we express it as
\begin{equation}
    E_{it} = \textstyle \sum_{m=1}^{M_{i}} \sum_t^{t + w_{imbt}}f(U_{mt}) 
    \label{energy_calculation}
\end{equation}
where $U_{mt}$ is the resource utilization ratio of MEC $m$ at time slot $t$. $f(*)$ is the power consumption model, describing the relationship between power cost and resource utilization.

To obtain the functional form of $f(*)$, we conduct a stress test on a Dell PowerEdge R760 bare-metal server equipped with two 14-core CPUs, as shown in Figure~\ref{powerstress}. During the test, we progressively increase the load using the \textit{stress} command. Meanwhile, the total server power consumption and the temperatures of both CPUs are continuously monitored through the Integrated Dell Remote Access Controller (iDRAC) system. The monitored data is retrieved remotely via the Simple Network Management Protocol (SNMP) using predefined Object Identifiers (OIDs) specific to CPU temperature and power consumption. These OIDs reside in the Management Information Base (MIB) provided by the manufacturer, which enables users to query the hardware performance metrics of the server. Based on the collected measurement data, a logarithmic curve fitting was performed using non-linear least squares regression, by minimizing the sum of squared residuals (equivalently minimizing the Mean Squared Error, MSE) between the measured data and the fitted logarithmic values, yielding the following model for server power consumption:
\begin{equation}
    f(U_{mt}) = 43.4779  \ln(100 \cdot U_{mt})+226.8324
    \label{server_power}
\end{equation}

\subsection{Problem formulation and dynamic network slicing}
The objective of this study is to maximize the long-term profits of multiple network slices by optimizing network resources to meet the dynamic demands of slices for latency and energy consumption while striving to limit re-optimization costs when the slice set changes.
Based on the aforementioned latency and energy consumption formulation, the optimization problem can be formulated as follows. For all constraints below, we have $\forall \psi \in \Psi$, $\forall t \in T_\psi$:

\begin{subequations}\label{eq:subeqns}

\begin{align}
P: & \max_{g_{im\psi t}, d_{il\psi t}}  \sum_{\psi \in \Psi}\sum_{t \in T_\psi}\sum_{i \in I_{\psi}} (\lambda_i W_{1\psi} / W_{i\psi t} + \rho_i E_{1\psi} / E_{i\psi t}) \label{6a} \\
\text{s.t. \ } & C1: 0 \leq U_{m\psi t} \leq 1,  \ \ \forall m \in M \label{6b} \\
&C2: 0 \leq g_{im\psi t}, \ \  \forall i \in I_{\psi}, \ \forall m \in M  \label{6c}\\
&C3: 0 \leq d_{il\psi t}, \ \  \forall i \in I_{\psi},\  \forall l \in L  \label{6d}\\
&C4: \textstyle 0 \leq \sum_{i\in I_{\psi}} g_{im\psi t} \leq J_{m}^{'}, \ \  \forall m \in M \label{6e}\\
&C5:  \textstyle 0 \leq \sum_{i\in I_{\psi}} d_{il\psi t} \leq B_l^{'}, \ \ \forall l \in L \label{6f}
\end{align} 
\end{subequations}
where in~\ref{6a}, $\psi$ represents a specific state of the system lasting for a duration of $\tau_\psi$, during which the set of active slices $I_{\psi}$ remains constant. Over an arbitrarily large time horizon $T$, the system state $\psi$ evolves dynamically within the set $\Psi$, due to the arrival and departure of slices.
$W_{1\psi}$ and $E_{1\psi}$ represent the minimum values for overall latency and energy consumption with the network slice set $\psi$, respectively. They serve to normalize these metrics such that latency and energy consumption remain comparable in magnitude. In practice, these minimum values are obtained by computing the average of several observed minima over a recent time window of $W_{i\psi}$ and $E_{i\psi}$. It is designed to mitigate the outliers and remove their side effect on the stability of the algorithm.
The weights $\lambda_i$ and $\rho_i$ represent the relative importance of latency and energy consumption preferences for slice $i$, with the constraint $\lambda_i + \rho_i = 1$. A network operator can derive these weights through a structured multi-attribute decision-making procedure \cite{wallenius2008multiple}. The process begins by specifying realistic best-to-worst ranges for both latency and energy consumption. Using an interactive weighting technique such as swing weighting, the operator then elicits managerial judgments on the relative desirability of improvements in each attribute, generating raw preference scores. Finally, these scores are normalized to sum to one, yielding the specific values of $\lambda_i$ and $\rho_i$ that guide subsequent resource-allocation decisions. In this study, these parameters are assumed to have been pre-established by the network operator.
Constraint $C1$ specifies that the utilization ratio of each MEC at any given time must not exceed 1. Constraints $C2$ and $C3$ guarantee that all slices receive allocations of computing and bandwidth resources on their respective servers and links. To avoid exceeding server and link capacities, Constraints $C4$ and $C5$ are introduced, where $J_m^{'}$ and $B_l^{'}$ represent the capacities of MEC $m$ and link $l$. These constraints lead to competition among slices for resources $g_{im\psi t}$ and $d_{il\psi t}$ while each slice strives to achieve its respective objective.
Consequently, the overall objective function is designed to balance the global performance across all slices. However, unlike traditional problems, the dynamic changes in the set $I_\psi$ continually update the problem settings, altering the interrelationships among existing slices. This leads to a shifting optimal solution space and diminishes the effectiveness of previously derived resource allocation strategies.

\begin{figure}[t]
    \centering
    \setlength{\abovecaptionskip}{0.3cm}
    \includegraphics[width=1\linewidth]{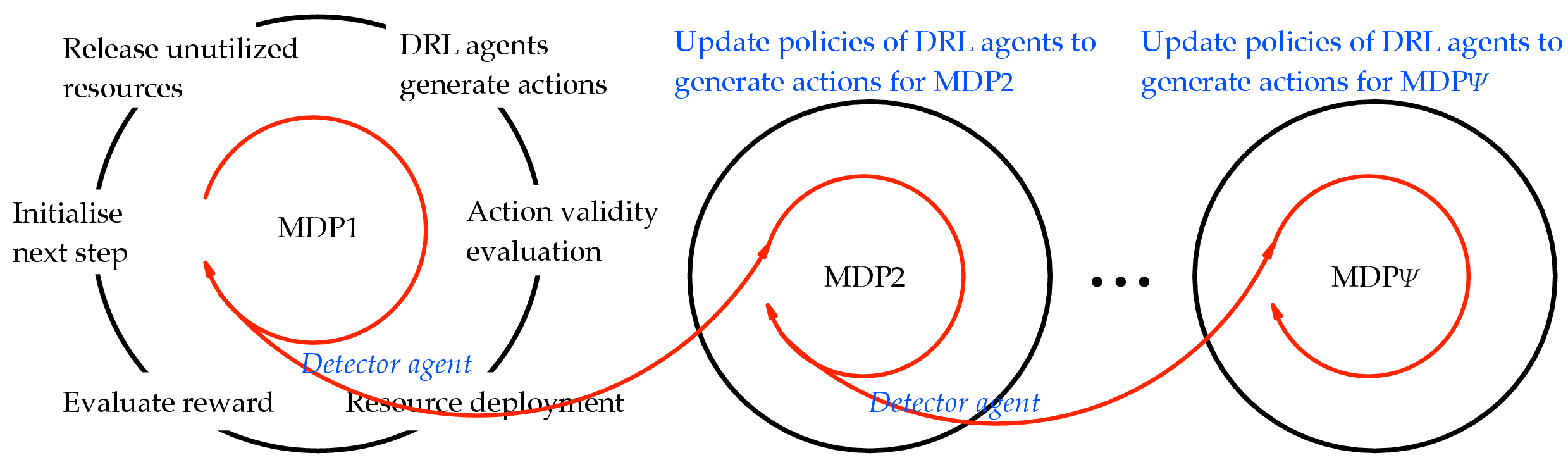}
    \vspace{-0.4cm}
    \caption{Iteration of dynamic DRL environment in MEC network resource management.}
    \label{drl_environment}
    \vspace{-0.4cm}
\end{figure}

\section{METHODOLOGY: Evolving MDP and Incremental MADDPG}
\label{sec:methodology}

\subsection{Evolving Markov Decision Process}
Considering the long-term resource dynamics in the network, we first formulate $P$ under a static $I_\psi$ as an MDP defined by the tuple $K = \langle S, A, H, R, \gamma\rangle$, in which $S, A, R, \gamma$ represent state space, action space, reward and discount factor (allowing to control the influence of future rewards) and ${H}$ represents the transition probability from any state $s_t\in {S}$ to any state $s_{t+1} \in S$ for any given action $a_{t} \in {A}$. 
In practical scenarios, however, the active slice set $I_\psi$ evolves over time. Such evolution results in changes to the original MDP tuple. Consequently, instead of a single stationary MDP, we model the network dynamics as a sequence of evolving MDP instances, each described by $\langle{S_{\psi}, A_{\psi}, H_{\psi}, R_{\psi}, \gamma_{\psi}} \rangle$ for $\psi \in \Psi$. Therefore, each problem instance $P_\psi$ entails its own state space, action set, horizon, and reward function, reflecting the temporal variations of network slices, as illustrated in Figure~\ref{drl_environment}.

\subsection{Multi-agent Deep Deterministic Policy Gradient}
In response, for each static MDP under the setup of $I_{\psi}$, we apply a fully cooperative multi-agent DDPG-based solution\footnote{The MADDPG methodology has been well-explained in a plethora of literature. Readers could refer to~\cite{tian2021multiagent, cheng2020joint, li2022drl} for more details. Unless otherwise specified, the following mentions of MADDPG refer to a fully cooperative setting.}, where each agent is assigned to a network slice to manage its resource allocation, and all agents collectively share a common reward. The architecture of the proposed solution is shown in Figure~\ref{fig:maddpg_architec}. The key elements of the MADDPG under $I_\psi$ are detailed as follows:

\begin{figure}[t]
    \centering
            \includegraphics[width=0.98\linewidth]{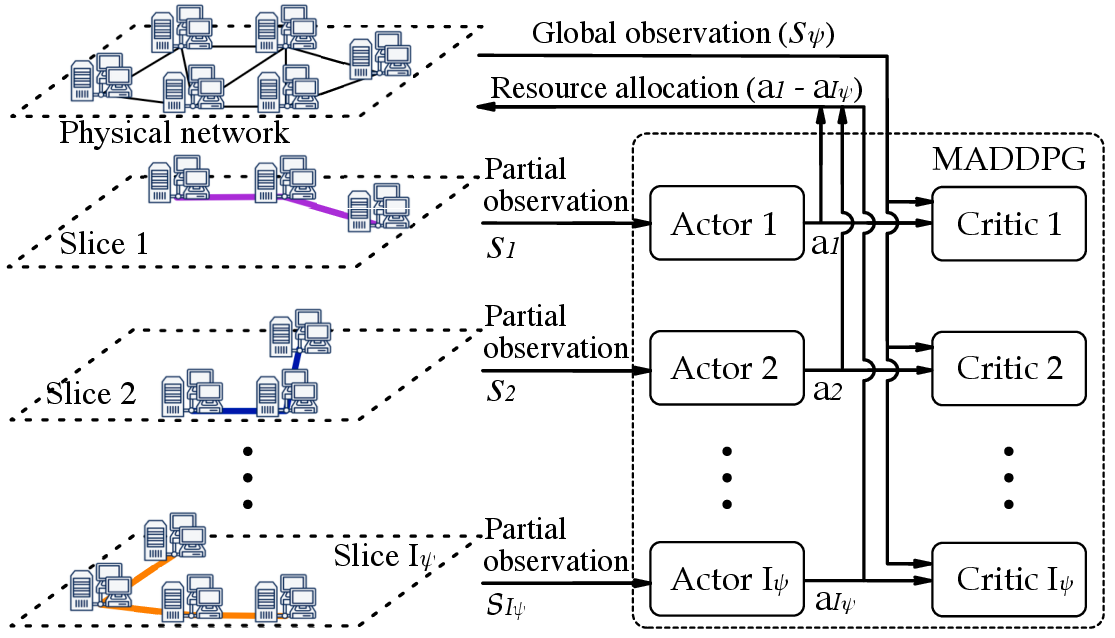}
    \caption{Architecture overview of the proposed MADDPG approach.}
    \label{fig:maddpg_architec}
\end{figure}

\textit{State of Actor and Critic $S_{\psi}$:} Each critic has a global view, accessing the resource information and the requests of the entire network. The state of a critic can be written as $s_{\psi t}$
\begin{equation}
s_{\psi t}= [\mathcal{J}_{M\psi t},  \mathcal{B}_{L\psi t}, \mathcal{G}_{I_\psi \psi t}, \mathcal{D}_{I_\psi \psi t}, \boldsymbol{\Lambda}_{\psi}] \ \ \ \forall t \in T_\psi, \psi \in \Psi
\label{critic_obs}
\end{equation}
where $\mathcal{J}_{M\psi t}$ and $\mathcal{B}_{L\psi t}$ represent the sets of remaining computing and bandwidth resources across all MEC servers and network links at time slot $t$ and state $\psi$. Specifically:
\begin{equation}
\mathcal{J}_{M\psi t}= \{J_m^{'}-\sum_{i \in I_\psi} g_{i m \psi t} \mid m \in M \}, \ t \in T_\psi, \psi \in \Psi
\end{equation}
\begin{equation}
\mathcal{B}_{L\psi t}= \{B_l^{'}-\sum_{i \in I_\psi} d_{i l \psi t} \mid l \in L \}, \ t \in T_\psi, \psi \in \Psi
\end{equation}
$\mathcal{G}_{I_\psi \psi t}$ and $\mathcal{D}_{I_\psi \psi t}$ represent computing resource requirements and workload size of all network slices at time slot $t$. $\boldsymbol{\Lambda}_{\psi} = \{(\lambda_i, \rho_i) | i\in I_\psi\}$ collects the latency- and energy-preference weights of all slices active in state $\psi$.
Conversely, the actor has a partial observation of the state $s_{it}$, which includes the remaining resource information of its utilized MECs and links, as well as the request information of the network slice. 
\begin{equation}
s_{i \psi t}= [J_{m\psi t }, B_{l\psi t}, G_{i\psi t}, D_{i\psi t}, \lambda_i, \rho_i]  \ \ \forall m \in M_{i},  l \in L_{i}
\label{actor_obs}
\end{equation}
For MEC without a request, related parameters are set to 0. In addition, it is worth noticing that the \textit{time slot} concept in MDP refers to a discrete decision point in the modeled environment. To resolve $P$, it is consistent with the time interval for processing requests by the network slice operator.

\textit{Action of Actor $A_{\psi}$:} Action $a_{i\psi t}$ can be represented as 
\begin{equation}
 a_{i\psi t} = [ \kappa a_{im\psi t} J_m^{'}, \forall m \in M_{i}; \kappa  a_{il\psi t} B_l^{'}, \forall l \in L_{i}]
\label{action}
\end{equation}
therefore $g_{im\psi t} = \kappa  a_{im\psi t}  J_m^{'}$, and $d_{il\psi t } = \kappa  a_{il\psi t}  B_l^{'}$. $a_{im\psi t}$ and $a_{i\psi tl}$ denote the allocation ratios for computing and bandwidth resources of MECs and links used by network slice $i$, respectively. $\kappa$ is used to constrain the maximum resources that can be allocated to each slice to avoid resource over-allocation. 
These ratios are determined by 
\begin{equation}
    a_{im\psi t} =clip(\tilde a_{im\psi t}+\epsilon \mathcal{O}(\mu, \sigma, \beta),0,1)
\label{OU1}
\end{equation}
\begin{equation}
    a_{il\psi t} =clip(\tilde a_{il\psi t}+\epsilon \mathcal{O}(\mu, \sigma, \beta),0,1)
\label{OU2}
\end{equation}
where $\tilde{a}_{im\psi t}$ and $\tilde{a}_{il\psi t}$ represent the original action outputs generated by the actor neural networks, set to $sigmod$ with an output range between $0\sim 1$.
$\mathcal{O}$ is the Ornstein–Uhlenbeck (OU) noise, and $\epsilon$ is the noise scale. $\mu$, $\sigma$, and $\beta$ are the mean, standard deviation, and speed of mean reversion of $\mathcal{O}$. The addition of this noise is to encourage action exploration, thereby enhancing the balance between exploration and exploitation and improving convergence stability. The clipping function is applied to ensure that the results remain within the valid range $[0, 1]$, preventing invalid network management option commands caused by noise. Moreover, a self-protection mechanism safeguards resource stability. If the requested resources remain within the available capacity, the full amount is deployed; otherwise, the system allocates only the residual capacity minus a 5\% safety margin, preserving a minimum buffer for critical operations.

\textit{Reward $R_\psi$:} Consistent with the objective function of formula~(\ref{6a}), the reward generated by each actor is set to 
\begin{equation}
r_{i\psi t} = ( \lambda_i W_{1\psi} / W_{i\psi t} + \rho_i E_{1\psi} / E_{i\psi t}) / I_\psi
\label{r1}
\end{equation}
The action \( a_{i\psi t} \) changes the network resource allocations \( J_{m\psi t} \) and \( B_{l\psi t} \), $\forall m \in M_{i},  l \in L_{i}$, resulting in the variation of next state \( s_{i\psi (t + 1)} \), and further determining \( W_{i\psi t} \) and \( E_{i\psi t} \) according to Equations~\eqref{3}, \eqref{4}, and \eqref{energy_calculation}, respectively.
In addition, the resource allocation strategies are formulated simultaneously by all actors based on state information, whereas the deployment of these strategies is executed sequentially within the network. During this process, if any MEC or link used by a slice runs out of resources, the network cannot serve the request. In such cases, the reward of this agent is set to 0. By aggregating the rewards from all agents, the shared reward can be written as
\begin{equation}
    r_{\psi t} = \sum_i^{I_\psi} r_{i\psi t}
    \label{r3}
\end{equation}
where $I_\psi$ in the denominator is used to normalize $r_{\psi t}$.
In network environments characterized by dynamic slicing requirements, the cooperative multi-agent solution offers several key advantages over single-agent and self-interested multi-agent approaches. First, resource-sharing introduces dependencies whereby the allocation for one slice can directly affect the performance of others. A multi-agent system, in which agents operate in parallel, adapts more effectively to these inter-slice interactions. Moreover, fully cooperative mechanisms emphasize overall performance, contrasting with self-interested approaches that aim to maximize individual gains without considering collective rewards~\cite{gronauer2022multi}.
Second, when each slice has distinct preferences for processing latency and energy consumption, a multi-agent framework enables each agent to focus on the specific requirements of its assigned slice. In contrast, a single-agent approach that treats the resource management of all slices as a unified problem often struggles to address the unique needs and interdependencies of individual slices.
Third, compared to single-agent solutions, multi-agent methods allow flexible trade-offs between the number of agents and the size of the action space. Lastly, multi-agent frameworks offer superior scalability relative to centralized single-agent solutions, enabling the system to adapt to variations in the number of slices by adding or removing agents without imposing changes to the action space.

\subsection{Delayed reward}
Different from conventional DRL approaches, where the environment immediately reveals the reward for each action, reward determination in practical slicing scenarios occurs only upon task completion. While a slice is being processed, the network operator may (i) receive new slice requests and allocate any remaining resources, or (ii) release resources from completed tasks. These ongoing allocations and deallocations result in fluctuating CPU utilization and, consequently, overall energy consumption. 
As a result, in addition to each action directly influencing future states and rewards as in a standard MDP, subsequent actions also retroactively affect the ultimate outcome of earlier actions due to the continuous evolution of the environment throughout the slice’s lifetime.
In response, we propose a delayed reward mechanism. In particular, transitions with unknown rewards are marked with an anchor. Once the reward becomes available, upon completion of the slice, it is merged with the corresponding transition using the anchor as a reference and stored in the replay buffer for subsequent training. This procedure enables the learning algorithm to precisely associate past and future actions with their final outcomes, incorporating their combined impact on energy consumption rather than relying on partial or premature estimates.

It is worth noting that this system does not constitute a Partially Observed MDP (POMDP) \cite{bennett2024proximal}. In a POMDP, the agent lacks full knowledge of the environment state and has to rely on belief states. By contrast, our agent has complete visibility of resource usage and network conditions. Moreover, the delayed-reward setting does not alter the fundamental DRL optimization principle. In specific, although the reward is deferred until slice completion, it is still uniquely determined by the trajectory of states and actions. Furthermore, 
in standard DRL, the influence of each action on cumulative rewards $R_t$ is implicitly captured through the discount factor $\gamma$. 
\begin{equation}
R_t = \sum_{k=t}^\infty \gamma^{k - t} R(s_{k}, a_{1k}, \ldots, a_{Ik})
\end{equation}
In the delayed-reward setting, subsequent actions build upon this foundation by also directly influencing the ultimate reward of earlier actions. This does not alter the objective of maximizing the cumulative discounted return.

With the collected buffer $\mathcal{D}$, the loss function for the critic update can be written as
\begin{equation}
V\left(\phi_i\right)=
\mathbb{E}_{s_t, a_{it}, r_{it}, s_{t+1} \sim \mathcal{D}}(\left(Q_{\phi_i}\left(s_t, a_{1t}, \ldots, a_{I t}\right)-y_{it}\right)^2)
\label{loss}
\end{equation}
where $y_{it}$ is the target Q-value. This target Q-value is updated using the Bellman equation~\cite{ding2020introduction} and can be expressed as:
\begin{equation} 
y_{it} = r_{t} + \gamma Q_{\phi_i}(s_{t+1}, a_{i(t+1)}, \ldots, a_{I(t+1)}) 
\label{targetQ}
\end{equation} 
This formula incorporates the immediate reward and the discounted future rewards captured by $Q_{\phi_i}$. In this context, the focus of \textit{long-term} is on capturing the cumulative impact of resource allocation decisions.

The gradient of the expected return $J$ and the update rule for the actor can be expressed as
\begin{equation}
\nabla_{\theta_i} J\left(\theta_i\right)=\mathbb{E}_{s_t \sim \mathcal{D}}(\left.\nabla_{\theta_i} Q_{\phi_i}\left(s_t, a_{1t}, \ldots, a_{I t}\right)\right|_{a_{jt}=\pi(s_t; \theta_j )})
\label{gradient}
\end{equation}
where $\pi$ refers to the policy of an actor agent and $j$ stands for all the agents in $I$.

\begin{algorithm}[t]
\caption{Incremental MADDPG with Delayed Rewards for Dynamic Network Slicing}
\label{alg:inc_maddpg}
\begin{algorithmic}[1]
    \PhaseLine{Initialization}
    \vspace{-0.2cm}
    \State Initialize actors $\pi_i$ with parameters $\theta_i$ and critics $Q_i$ with parameters $\phi_i$, their target networks $(\pi_i^-, Q_i^-)$ using $(\theta_i^-, \phi_i^-)$, replay buffer $\mathcal{D}$, and anchor map $\mathcal{A}$ for delayed rewards.

    \PhaseLine{Training on Slice Set $I_\psi$}
    \vspace{-0.2cm}
    \For{$t=0$ to $T_\psi-1$}
        \ForAll{$i\in I_\psi$ \textbf{(in parallel)}}
            \State Observe $s_{\psi t}$ and $s_{i\psi t}$.
            \LongState{Actor compute raw ratios
            $\tilde a_{im\psi t},\tilde a_{il\psi t} = \operatorname{sigmoid}(\pi_i(s_{i\psi t};\theta_i))$.}
            \LongState{Add OU exploration and clip via Eq.~\eqref{OU1}, \eqref{OU2}.}
        \EndFor
            \State{Execute joint action; observe next states $\{s_{i\psi(t+1)}\}$.}
        \If{reward available at $t$}
            \LongState{Compute rewards using Eq.~\eqref{r1}, \eqref{r3}} 
            \State Push transition $(s_t,a_t,r_t,s_{t+1})$ into $\mathcal{D}$.
        \Else
            \LongState{Create anchor\_id and store incomplete transition in $\mathcal{A}$.}
            \LongState{On task completion, compute the reward via Eq.~\eqref{r1}, \eqref{r3} and merge by anchor\_id into $\mathcal{D}$.}
        \EndIf
        \If{update due}
            \LongState{Sample a mini-batch from $\mathcal{D}$; update actor \& critic via Eq.~\eqref{loss}, \eqref{targetQ} and \eqref{gradient}.}
            \LongState{Soft-update targets: $\theta_i^- \!\leftarrow\! \tau\theta_i+(1{-}\tau)\theta_i^-$, \ \ $\phi_i^- \!\leftarrow\! \tau\phi_i+(1{-}\tau)\phi_i^-$.}
        \EndIf
    \EndFor

    
    \PhaseLine{Incremental Update on $I_\psi \!\rightarrow\! I_{\psi+1}$}
    \vspace{-0.2cm}
    \If{slice set changes}
        \If{new slice $n$ appears}
            \LongState{Compute weights $\{\eta_i\}$ by Eq.~\eqref{incre1}; initialize the actor and critic of new slice $n$ via Eq.~\eqref{incre2}, \eqref{incre3}.}
        \ElsIf{some slices are removed}
            \LongState{Form a generalized model via Eq.~\eqref{incre1}--\eqref{incre3} and assign to remaining agents.}
        \EndIf
        \State Resume \textbf{Phase 2} with updated $I_{\psi+1}$.
    \EndIf
\end{algorithmic}
\end{algorithm}

\subsection{Incremental learning}
Although MADDPG is capable of solving a static optimization problem, the trained model may lose effectiveness when the slice set changes. To enhance the generalization capability of the model, three strategies are commonly employed: transfer learning~\cite{weiss2016survey}, meta-learning~\cite{finn2017model}, and incremental learning (also referred to as lifelong or online learning). While these methods share similar foundational principles, they each have distinct objectives suited to various contexts.
Specifically, transfer learning involves adapting knowledge from a source task to a related target task. Although the source and target tasks differ, they share underlying similarities, such as using a cat recognition model to assist in recognizing dogs. This approach is typically applied in offline neural network training, where a model is trained once on a source dataset and then fine-tuned or applied to a different target dataset. Meta-learning, by contrast, aims to “learn how to learn.” It trains a model on a diverse set of tasks to enhance its ability to quickly learn new tasks. Generally an offline training approach, meta-learning focuses on building flexibility across varied tasks rather than optimizing performance on a single type of data.
In comparison, incremental learning is designed for online scenarios, enabling continuous model adaptation in dynamic environments. By updating its knowledge progressively, a model trained with incremental learning can continuously adapt to new data, extending beyond offline adjustments and into real-time learning. 
Given the online training nature of DRL and the dynamic characteristics of network environments within a single-task setting, we employ incremental learning to minimize re-optimization costs. This choice allows the proposed MADDPG to proactively adapt to changes in the slice set $I_\psi$ while retaining previously learned decision-making knowledge.

A proximity-weighted strategy is employed to obtain the policy for the new agent model. This involves comparing new slices with existing ones in terms of their preferences for latency and energy consumption, denoted as $\lambda_n$ and $\rho_n$, respectively. Agents of slices with similar requirements will receive greater weight in the inheritance of policies. This approach can be written by

\begin{equation}
\eta_{i} = \frac{\frac{1}{\left|\lambda_i-\lambda_n\right|+\left|\rho_i-\rho_n\right|} } {\sum_i^I \frac{1}{\left|\lambda_i-\lambda_n\right|+\left|\rho_i-\rho_n\right|}} 
\label{incre1}
\end{equation}
\begin{equation}
\phi_{\psi + 1} = \sum_i \eta_i \phi_{i\psi}
\label{incre2}
\end{equation}
\begin{equation}
\theta_{\psi + 1} = \sum_i\eta_i \theta_{i\psi}
\label{incre3}
\end{equation}
where $\phi_{i\psi}$ and $\theta_{i\psi}$ represent the parameters of the critic and actor networks of $i$. To mitigate the issue of division by zero arising from uniformity in preferences, the minimum $\left|\lambda_j-\lambda_n\right|$ and $\left|\rho_j-\rho_n\right|$ are set to 0.1.
When the number of agents increases, the generalized model is loaded into the new agent while preserving the models of the existing agents. In the opposite case, where the number of agents decreases (i.e., decremental learning), the newly derived model is assigned to all agents. 
Note that, in practical scenarios, policy inheritance could potentially benefit from considering not only the slices’ latency/energy preferences but also their traffic load statistical characteristics.
In the experiments, however, slice traffic is randomly generated without assuming specific statistical similarity. Consequently, load-based similarity is not included in the inheritance weight calculation in Equation~\ref{incre1}.


\begin{figure}[t]
    \centering
            \includegraphics[width=0.96\linewidth]{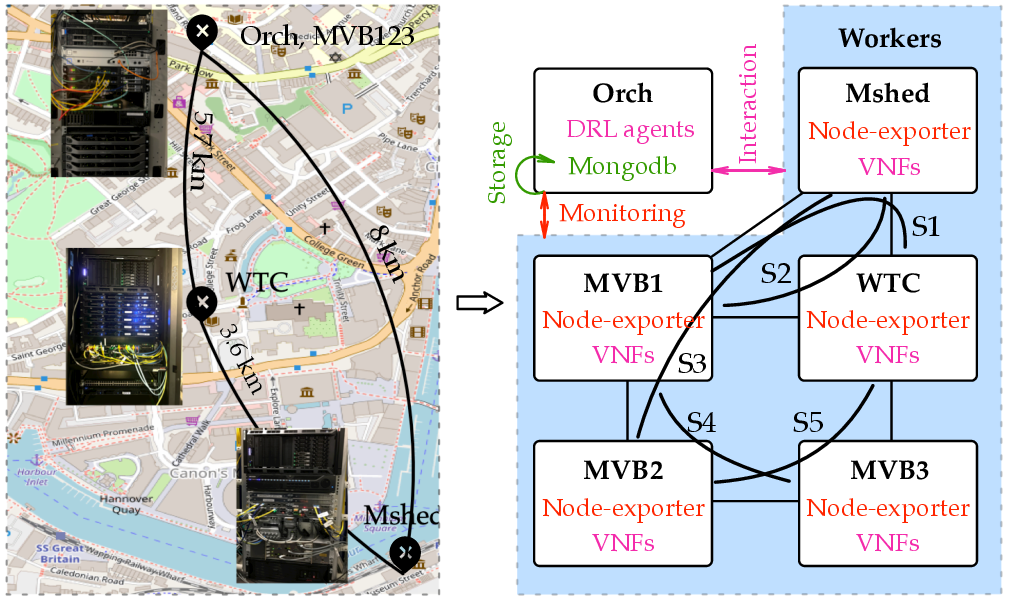}
    \caption{City-wide urban testbed spanning edge facilities at MVB, WTC, and MShed. Green, red, and purple indicators represent the deployment of storage, monitoring, and network slice execution elements across the nodes, respectively.}
    \label{testbed_1}
\end{figure}

\section{Validation over an urban testbed}
\label{sec:evaluation}
To validate the performance of our solution, we construct an experimental testbed consisting of six Virtual Machines (VMs) deployed across three locations in Bristol, which are then used for subsequent trials. The following section first presents the testbed setup, including how the DRL framework monitors and controls the network, and then discusses the results of our proposed algorithm.

\begin{figure}[t]
    \centering
            \includegraphics[width=1\linewidth]{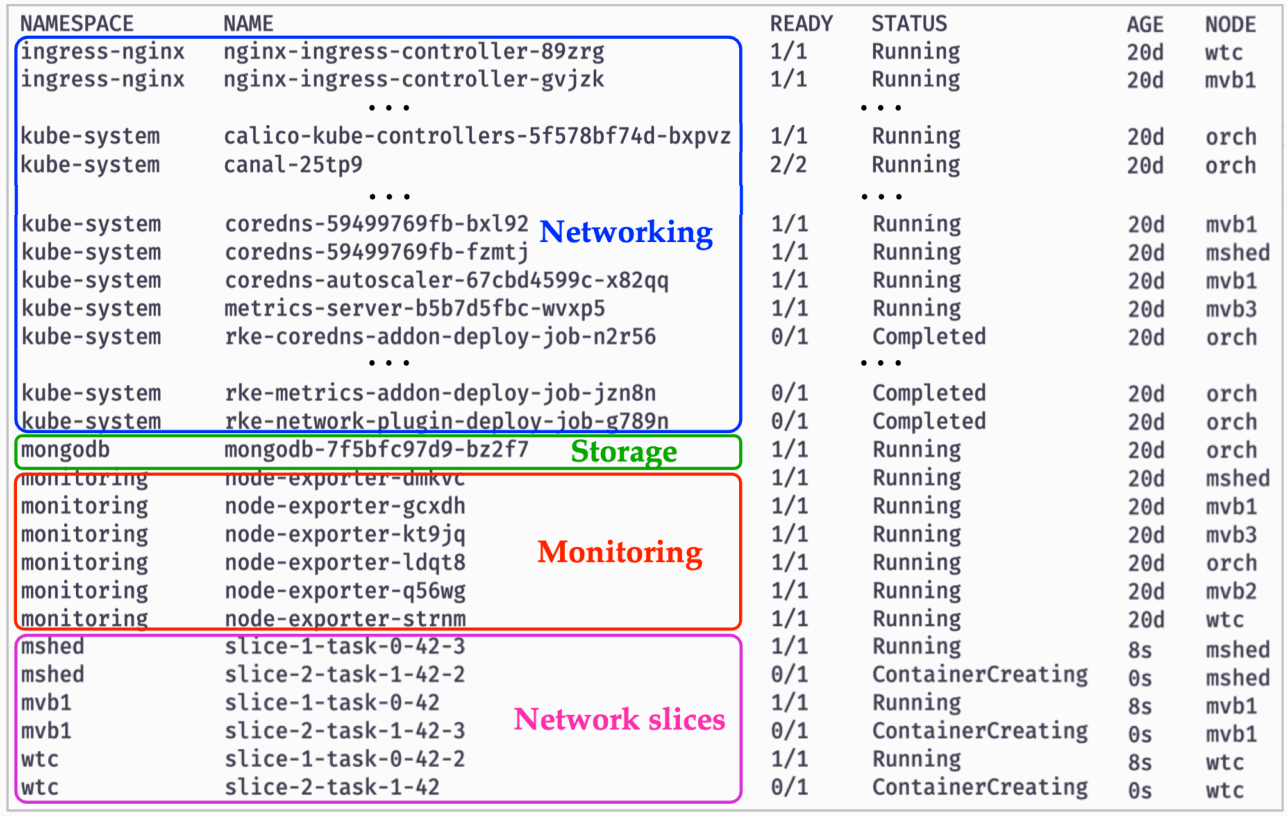}
    \vspace{-0.4cm}
    \caption{A snapshot of the network status. Blue, green, red, and purple indicators denote different pods responsible for networking, storage, monitoring, and network slice execution, respectively.}
    \vspace{-0.1cm}
    \label{testbed_realtime}
\end{figure}

\begin{table}[t]
  \centering
  \caption{Performance comparison of various resource management solutions within a 4-slice edge network scenario in the testbed.}
  \vspace{-0.2cm}
    \begin{tabular}{lllr}
\cmidrule[0.7pt]{1-3}    \multicolumn{1}{p{8em}}{Solutions} & \multicolumn{1}{p{10em}}{Average energy (kJ) \newline{}\& variance} & \multicolumn{1}{p{9em}}{Average latency (s) \newline{}\&  variance} &  \\
\cmidrule{1-3}    Random & 17.5794 \& 1.1397 & 15.1684 \& 0.9601 &  \\
    Over allocation & 7.9252 \& 0.0227 & 7.5636 \& 0.3694 &  \\
    Static & 29.9133 \& 0.7995 & 21.0051 \& 0.3898 &  \\
    SI-MADDPG & 17.1522 \& 0.1672 & 14.9122 \& 0.1428 &  \\
    FC-MADDPG & 16.4348 \&  0.1012 & 15.0280 \& 0.1144 &  \\
\cmidrule{1-3}    \multicolumn{1}{p{8em}}{Solutions} & \multicolumn{1}{p{10em}}{Objective function value  \newline{}\&  variance} & \multicolumn{1}{p{9em}}{Slice success\newline{}service rate} &  \\
\cmidrule{1-3}    Random & 0.2494 \& 0.0643 & 96.80\% &  \\
    Over allocation & 0.3455 \& 0.1024 & 50\%  &  \\
    Static & 0.2544 \&  0.0320 & 96.10\% &  \\
    SI-MADDPG & 0.6408 \& 0.0005 & 100\% &  \\
    FC-MADDPG & 0.6698 \& 0.0004 & 100\% &  \\
\cmidrule[0.7pt]{1-3}    \end{tabular}%
  \label{testbed_result_1}%
\end{table}%

\begin{figure*}[!t] 
    \centering 
    \setlength{\subfigcapskip}{-0.2cm}
    \subfigure[Processing latency of FC-MADDPG]{
        \label{cdf1}
                    \includegraphics[width=0.45\linewidth]{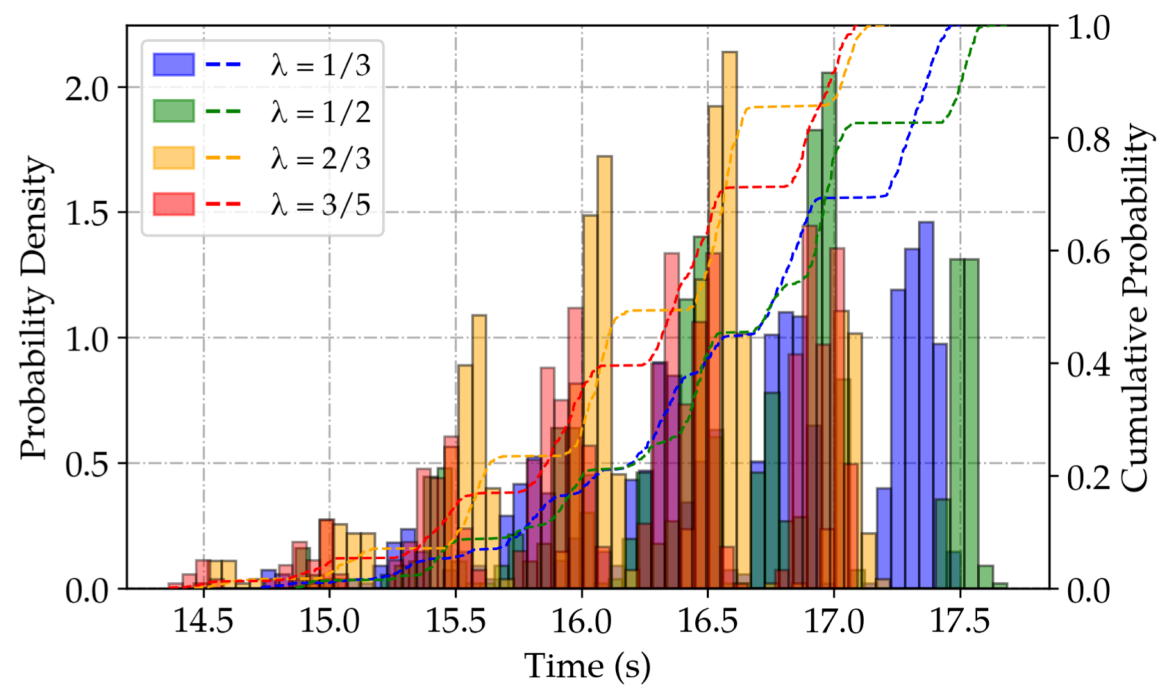}}
    \subfigure[Energy consumption of FC-MADDPG]{
        \label{cdf2}
                    \includegraphics[width=0.45\linewidth]{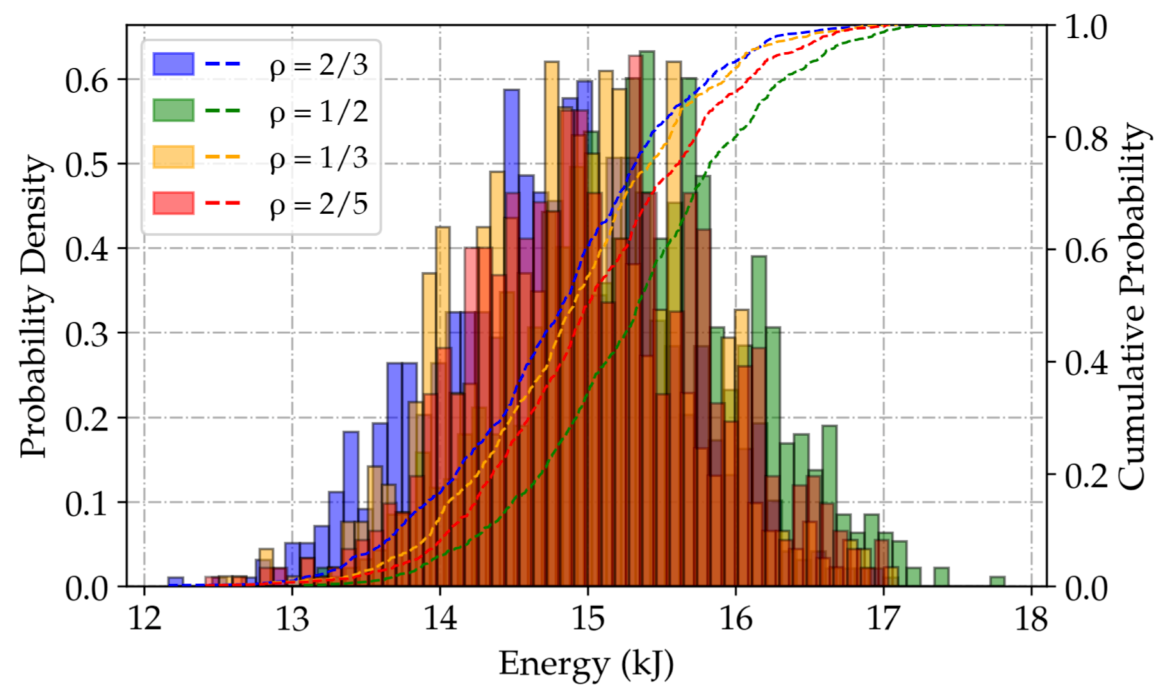}}
        \subfigure[Processing latency of SI-MADDPG]{
        \label{cdf3}
                    \includegraphics[width=0.45\linewidth]{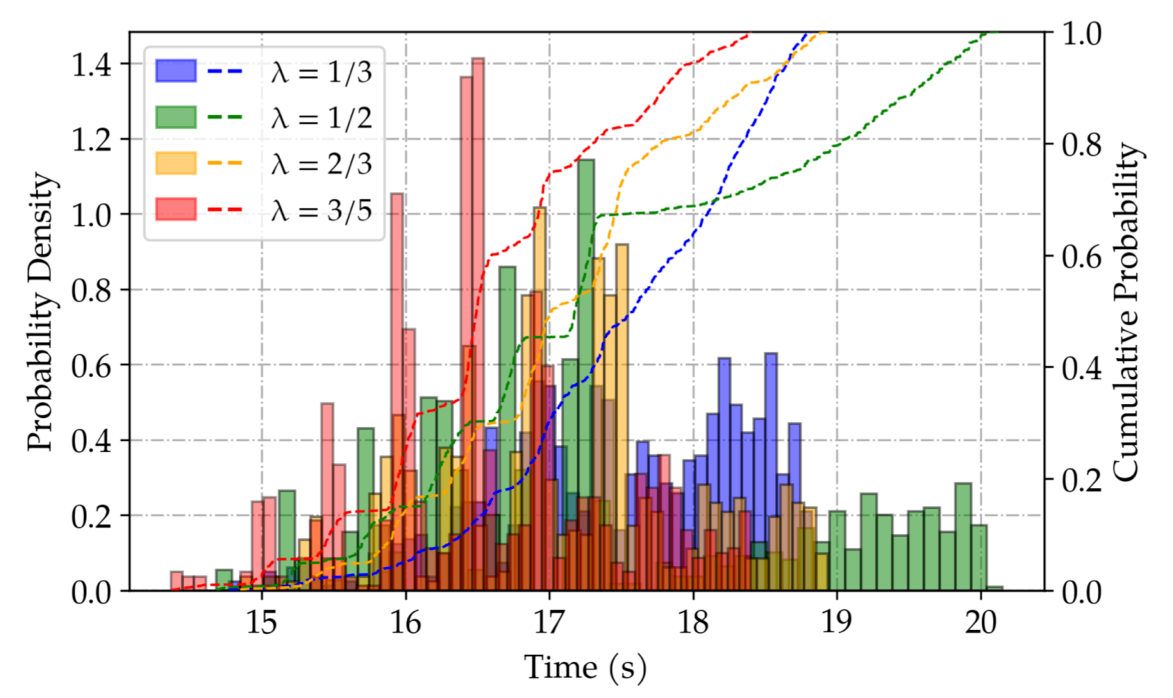}}
    \subfigure[Energy consumption of SI-MADDPG]{
        \label{cdf4}
                    \includegraphics[width=0.45\linewidth]{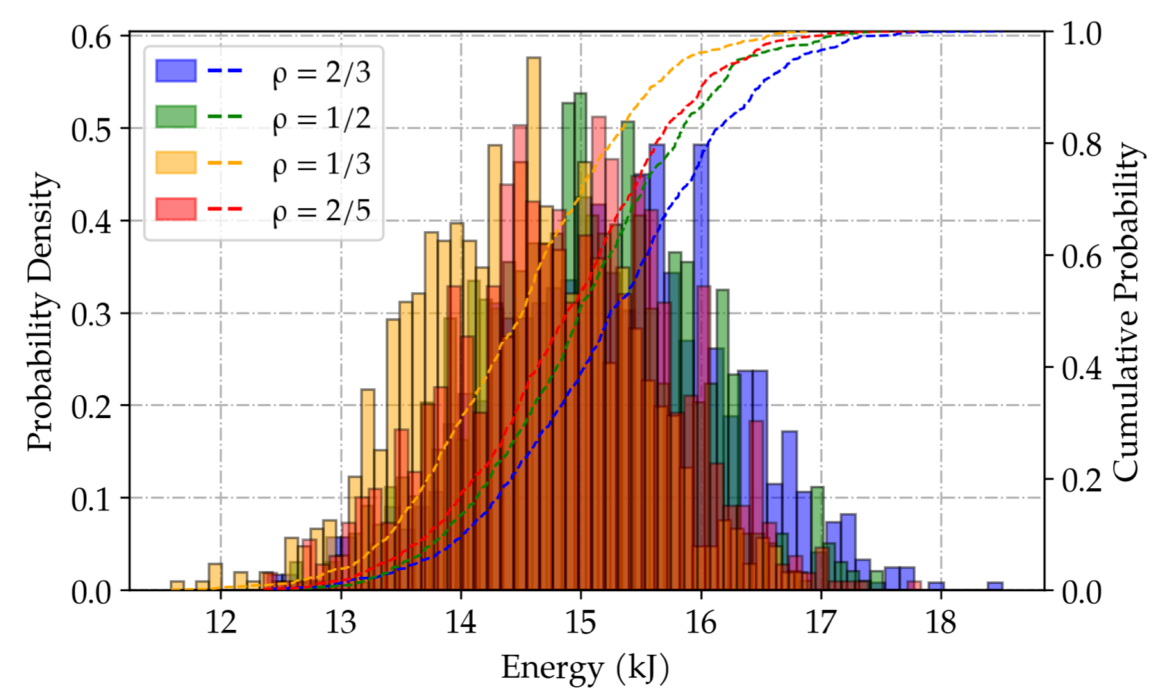}}
    \vspace{-0.2cm}
    \caption{PDF and CDF of processing latency and energy consumption for slices with different preferences managed by FC-MADDPG and SI-MADDPG based on 1,000 inference results.}
    \label{cdf}
\end{figure*}

\vspace{-0.3cm}
\subsection{Testbed setup \& DRL deployment}
\label{sec:setup}

\subsubsection{Testbed setup}
The testbed architecture is shown in Figure~\ref {testbed_1}. It consists of three sets of physical servers distributed across the Merchant Venturers Building (MVB), the MShed Museum, and the We The Curious Museum (WTC). Using OpenStack, we deployed six VMs across these servers, each equipped with 4 CPU cores to provide computing resources. i.e $J_{m}^{'} =4, \ \forall  m \in M$. Additionally, a 10 Gbps inter-node bandwidth limitation, $B_l^{'}$, is enforced by \textit{tc} command across all connections $L$. The network processes all incoming requests at 10-second intervals. Slice requests are emulated by a Python script. At each decision interval, their workload $D_{il\psi t}$ and CPU demands $G_{im\psi t}$ are randomly drawn from 25–35 Gb and 10.8–13.2 CPU-cores, respectively.

\subsubsection{Action deployment}
Among the six VMs, a Kubernetes cluster is deployed, where each node is a VM and each VNF component in a slice can be considered a Kubernetes pod. 
Based on the DRL action, the Kubernetes API is invoked to create the required pods equipped with \textit{traffic control (tc)} to enforce their CPU and bandwidth allocations. The allocated resources are then saturated with the \textit{stress} and \textit{iperf3} commands. $\kappa$ is set to 40\%.

\subsubsection{State monitoring}
To extract the state information from the network, the network request related variable sets 
$\mathcal{G}_{I_\psi\psi t}$, $\mathcal{D}_{I_\psi\psi  t}$ and $\boldsymbol{\Lambda}_{\psi}$ can be directly retrieved from the emulator. A \textit{Prometheus node-exporter} is deployed as a \textit{DaemonSet} on all worker nodes, monitoring the computing and bandwidth resource information, thereby collecting data for variable sets $\mathcal{J}_{M\psi t}$ and $\mathcal{B}_{L\psi t}$. The monitoring process is executed at one-second intervals to capture rapid CPU usage variations.
On the control node, concurrent HTTP GET requests are dispatched to the \textit{Node exporter} RESTful API endpoints of all servers to retrieve the monitored network information, which is then stored in a MongoDB database.
During training or inference, the actor and critic models allocated in the Orch can access the network resource data from the MongoDB collection using the specified IP address and pre-defined collection name.

Overall, these network components are demonstrated in a snapshot as shown in Figure~\ref{testbed_realtime}. Since both (i) action execution via the Kubernetes API and (ii) state/reward monitoring via \textit{Node exporter} are not confined to the specific experimental settings adopted in this study, the proposed solution is portable to different networks with diverse topologies, resource allocations, or service demands.

\subsection{Experimental results}

Before exploring the performance of incremental learning of the MADDPG, it is crucial to verify the baseline effectiveness of the algorithm.
Table~\ref{testbed_result_1} summarizes the objective function value, average energy consumption, average latency, and success service rate of various network slice management solutions in a 4-slice scenario (S1\,-\,S4). Besides the proposed fully cooperative MADDPG (FC-MADDPG), we include a random allocation strategy, which distributes resources randomly among slices without considering current load, task requirements, and network conditions; an over allocation strategy, which allocates max available resources to each slice at each time slot.
A static slicing-based solution proposed by~\cite{caballero2019network}, which refers to a partitioning of resources based on the requirement ratio is also accessed, as well as a self-interested MADDPG, where each agent optimizes its performance without considering others.
A single-agent DDPG (SA-DDPG) approach is also tested, where an actor network is applied to determine resource allocation across all slices at each time step while it is not presented because repeated training attempts failed to achieve convergence, highlighting the advantage of the multi-agent setting in improving training stability and facilitating convergence. $\kappa$ is applied across all solutions to avoid resource over-allocation and increase the success service rate. 

Table~\ref{testbed_result_1} indicates that both DRL-based schemes surpass the random and static baselines on latency and energy, achieving lower means and smaller variances. In addition, while over-allocation attains the minimum delay and power draw, its service success rate drops to 50\% because over-provisioning exhausts resources needed for later arrivals. Between the two DRL variants, energy expenditure is almost the same, while the Fully Cooperative (FC)-based solution delivers shorter latency, demonstrating better holistic optimization capability. To assess alignment with slice-specific preferences, Figure~\ref{cdf} presents the Probability Density Function (PDF) and Cumulative Distribution Function (CDF) of processing latency and energy consumption of the two DRL-based solutions for four slices with distinct latency and energy preferences based on 1,000 inference results. Both methods largely respect these priorities; for instance, slices with $\lambda=2/3$ and $3/5$ exhibit lower and more concentrated processing latency distributions, demonstrating superior performance for latency-sensitive tasks. Nonetheless, FC-MADDPG achieves a superior latency–energy balance for the energy-oriented slice ($\rho$ = 2/3), as shown by the comparison between Figures \ref{cdf2} and \ref{cdf4}, whereas SI-MADDPG incurs the highest energy cost in that case.

\begin{figure}[t] 
    \centering  
            \includegraphics[width=0.85\linewidth]{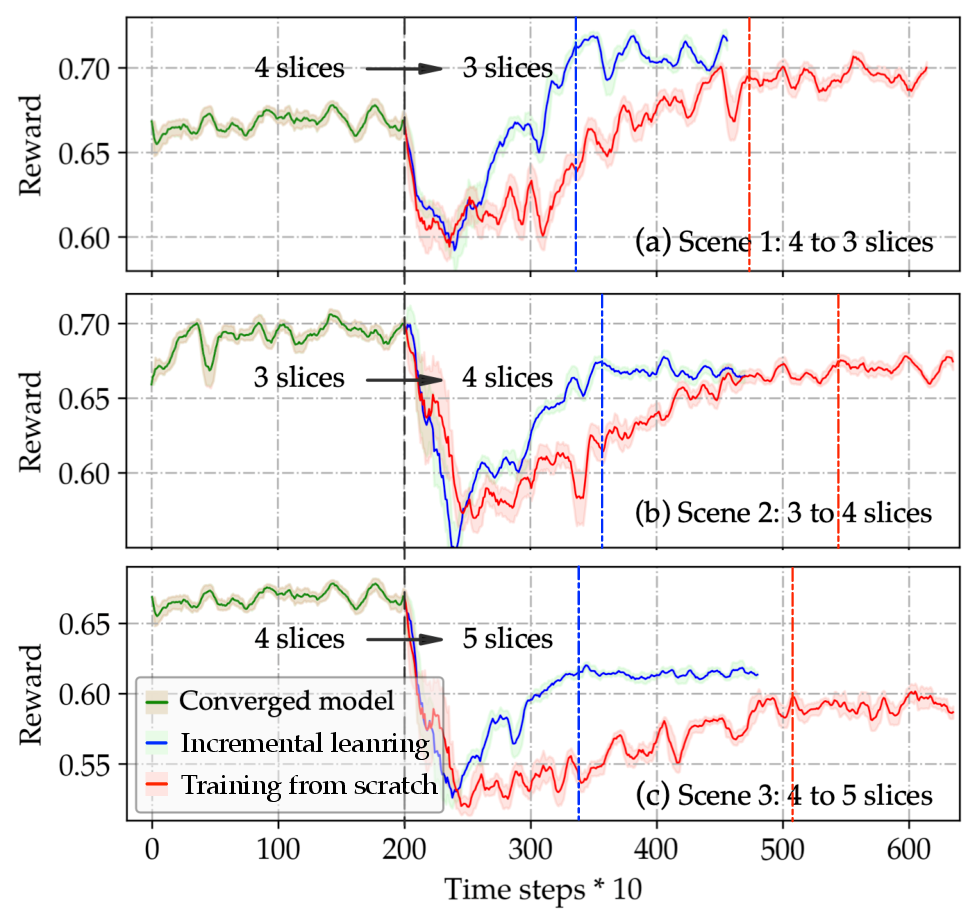}
    \vspace{-0.2cm}
    \caption{Performance evaluation of the proposed incremental MADDPG with varying network slices in the testbed. The green line represents the inference process of the converged MADDPG in a certain network slice setup ($I_\psi$). Blue and Red demonstrate the incremental learning and training from scratch process of the MADDPG under a new slice set ($I_{\psi + 1}$). Notice the differences in the Y-axis scale between the graphs applied to enhance the readability of results.}
    \label{incremental_2}
    \vspace{-0.2cm}
\end{figure}

\begin{figure}[t] 
    \centering  
            \includegraphics[width=0.96\linewidth]{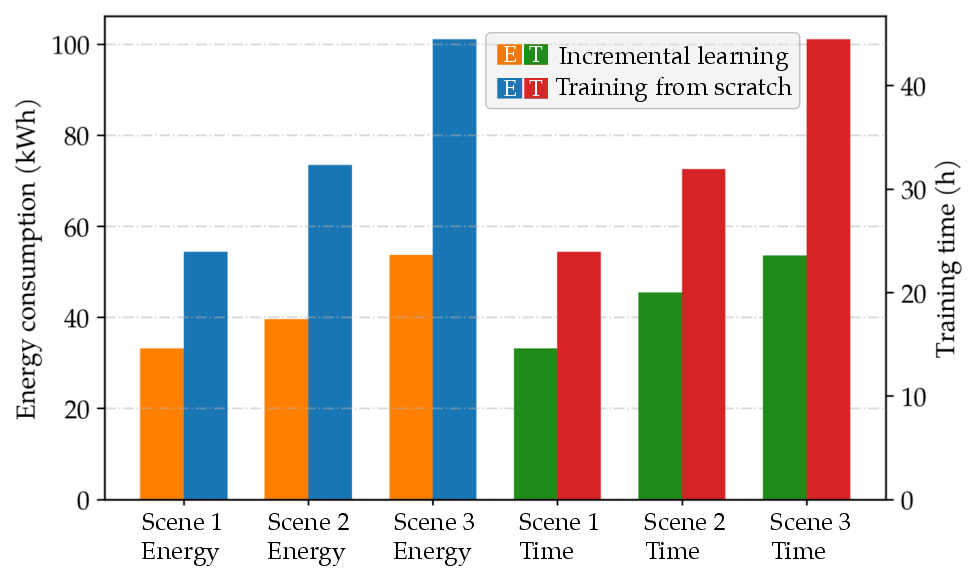}
    \vspace{-0.2cm}
    \caption{Energy consumption \& Training time comparison of MADDPG between incremental learning and training from scratch on the testbed.}
    \label{testbed_result_3}
\end{figure}

To further corroborate the effectiveness of the incremental MADDPG scheme, we conduct an additional set of experiments in which a policy that has already converged on a given MDP $\psi$ is required to adapt to environments with different sets of network slices. Specifically, we examine three transition scenes: 4 to 3 slices, 3 to 4 slices, and 4 to 5 slices. For each scenario, we compare the re-optimization behavior of (i) incremental learning, which inherits the previously converged policy, and (ii) training from scratch. As summarized in Figure \ref{incremental_2}, the proposed incremental learning solution demonstrates approximately a threefold reduction in re-optimization steps across various slice configuration changes.
Moreover, Figure~\ref{testbed_result_3} quantifies the energy and training time gains delivered by the incremental MADDPG across the three re-optimization scenes. Energy consumption is measured system-wide, including (i) the computational load of DRL model updates executed on the controller (Orch) node, and (ii) the operational overhead incurred on the worker nodes, where network slices must actively react to the DRL’s provisional actions throughout training. 
As the number of slices increases, the training time correspondingly rises because additional slices reduce the resources allocated to each slice, thereby prolonging processing time and exacerbating energy consumption.
Across the three comparisons, incremental learning consistently demonstrated significant advantages over training from scratch, achieving energy savings of approximately 40\% to 50\%. 

\subsection{Discussion} \label{sec:discussion}

\textit{Scalability:} The scalability of the proposed algorithm is determined by two main factors: the number of network slices and the size of the network. In the proposed setup, each slice is assigned an agent to simulate resource competition among slices. Therefore, if the number of slices increases, the number of agents will also increase. However, this does not affect the state and action space of each actor agent; it only impacts the state space of the critic agent due to the additional network request information. On the other hand, if the network size grows with more nodes and links, unlike the variation in the number of slices, the state information of both the actor and critic networks will increase to accommodate the additional computing and bandwidth resource information.

\textit{Stability:} The stability of the algorithm determines its feasibility in real-world applications. Therefore, the convergence rate of the MADDPG is tested.
More than 10 experiments are conducted for each slice configuration, encompassing both incremental learning and baseline scenarios, with non-convergence occurring in no more than 2 instances per scenario. Moreover, the final convergence objective function values are consistent across all test runs within each scenario, suggesting a high degree of stability of the strategy.

\textit{Generalization ability:} The lack of this ability is a common bottleneck for neural networks due to the input feature distribution drift. The proposed strategy is only applicable to variations in the number of slices and does not yet address the impacts of all the dynamic factors in 5G networks, such as changes in network topology caused by the addition or removal of servers. Future works can be conducted on these perspectives.

\textit{Limitations of abstract VNF modeling:} The current framework abstracts subscriber demands as generic VNFs, without explicitly modeling standard 5G functions such as the  AMF, SMF, or UPF. While this abstraction improves generality, it limits the ability to capture protocol-specific behaviors and latency characteristics introduced by individual network functions. In future works, we plan to incorporate explicit models of such functions to enable finer-grained latency analysis and improve the fidelity of the system model.


\section{CONCLUSION}
\label{sec:conclusion}

This paper explores the resource allocation and scheduling of network slices in edge computing networks. In the absence of effective resource management in network slicing, service failures and performance fluctuations become potential concerns. To address these issues, we proposed a MADDPG-based solution that allocates a DRL agent to manage resources on each network slice; therefore capturing the resource-sharing relationships among network slices. Additionally, the MADDPG algorithm is enhanced with incremental learning features to capture network slice dynamics and reduce re-optimization costs.
Experimental results, benchmarked against existing solutions, reveal substantial improvements achieved by MADDPG. In addition, the incremental learning aspect demonstrates robust generalization capability in capturing varying slice number dynamics, leading to substantial enhancements in training speed and cost reduction compared to pre-trained base models. Most importantly, our real-world implementation showcases the effectiveness of the proposed incremental MADDPG in live settings, highlighting its resilience against instabilities in network management processes.

\section*{ACKNOWLEDGMENT}
The authors would like to express their gratitude for the support from the UK-funded project REASON. This work was also partially supported by the EPSRC Future Telecoms Research Hub, Platform for Driving Ultimate Connectivity (TITAN) under Grant EP/X04047X/2 and Grant EP/Y037243/1.

\bibliographystyle{IEEEtran}\bibliography{references}

\vfill\pagebreak

\end{document}